\documentclass[conference,harvard,brazil,english]{sbatex}

\makeatletter
\def\verbatim@font{\normalfont\ttfamily\footnotesize}
\makeatother

\usepackage[utf8]{inputenc}

\usepackage{icomma} 
\usepackage{amsmath,amssymb,amsfonts} 
\usepackage{verbatim}
\usepackage[brazil,english]{babel}
\usepackage{graphicx} 
\usepackage{psfrag} 
\usepackage{float} 
\usepackage{algorithm}
\usepackage{algpseudocode}
\usepackage{multirow}
\usepackage{multicol, blindtext, graphicx}
\usepackage{color}
\usepackage{xcolor}
\usepackage{epstopdf}

\usepackage{url} 

\newdimen\slantmathcorr
\def\oversl#1{
    \setbox0=\hbox{$#1$}
    \slantmathcorr=\wd0
    \hskip 0.2\slantmathcorr \overline{\hbox to 0.8\wd0{%
            \vphantom{\hbox{$#1$}}}}
    \hskip-\wd0\hbox{$#1$}
}

\def\undersl#1{
    \setbox0=\hbox{$#1$}
    \slantmathcorr=\wd0
    \underline{\hbox to 0.8\wd0{%
            \vphantom{\hbox{$#1$}}}}
    \hskip-0.8\wd0\hbox{$#1$}
}

\begin{document}

\title{INTERVAL SIMULATION OF NARMAX MODELS BASED ON COMPUTER ARITHMETIC}

\author{Priscila F. S. Guedes}{pri12\_guedes@hotmail.com}
\address{GCOM - Grupo de Controle e Modelagem \\ UFSJ - Universidade Federal de São João del-Rei\\ Pça. Frei Orlando, 170 - Centro - 36307-352 - São João del-Rei, MG, Brasil}

\author[1]{Márcia L. C. Peixoto}{marciapeixoto93@hotmail.com}

\author[1]{Otávio A. R. O. Freitas}{otavio\_augusto\_freitas@hotmail.com}

\author[2]{Alípio M. Barbosa}{alipiomonteiro@yahoo.com.br}
\address{Centro Universitário Newton Paiva \\  Rua José Cláudio Rezende, 420 - Estoril - 30494-230 - Belo Horizonte, MG, Brasil}

\author[1]{Samir A. M. Martins}{martins@ufsj.edu.br}

\author[1]{Erivelton G. Nepomuceno}{nepomuceno@ufsj.edu.br}

\twocolumn[

\maketitle


\selectlanguage{english}
\begin{abstract}

System identification is an important area of science, which aims to describe the characteristics of the system, representing them by mathematical models. Since many of these models can be seen as recursive functions,  it is extremely important to control the errors in these functions, because small errors introduced in each computational step can grow exponentially due to the sensitivity to initial conditions present in this type of functions. One of the ways to control rounding and truncation errors is through interval arithmetic, since it is not possible to represent all numbers in the computer because of the finite representation in them. Thus, in arithmetic interval a number is represented by an interval in which the true number is within that interval. In this manuscript we developed an algorithm that performs the operations of interval arithmetic using basic functions. We have compared compared our results with the Matlab-toolbox Intlab. Numerical experiments have shown that our method is superior producing narrower intervals.

\end{abstract}


\selectlanguage{brazil}
\begin{abstract}

Identificação de sistemas é uma área importante da ciência, tendo como objetivo descrever as características do sistema, representando-as por equações matemáticas. Como muitas dessas equações podem ser vistas como funções recursivas,  é de extrema importância controlar os erros nessas funções, já que pequenos erros introduzidos a cada passo computacional pode crescer exponencialmente devido a sensibilidade a condições iniciais presente nessas funções. Uma das formas de controlar erros de arredondamento e truncamento é através da aritmética intervalar, uma vez que não é possível representar todos os números no computador devido à representação finita dos mesmos. Então, na aritmética intervalar, um número é representado por um intervalo em que o verdadeiro número se encontra nesse intervalo. Foi desenvolvido um algoritmo que realiza as operações da aritmética intervalar a partir de funções básicas do Matlab. Os resultados foram comparados com o toolbox do Matlab, Intlab. Experimentos numéricos mostraram que os intervalos obtidos pelo nosso métodos são menores do que aqueles obtidos no Intlab.

\end{abstract}

]

\selectlanguage{brazil}

\section{Introduction}
\selectlanguage{english}
\hyphenation{FAPEMIG NARMAX}

System identification plays a fundamental role in the estimation of mathematical models from data \cite{ljung1999}. The estimated models can then be used to analyse input-output relationships, to simulate the system in various situations and for controller design \cite{BTA2015}.

\begin{figure}[ht!]
    \centering
    \includegraphics[width=0.47\textwidth]{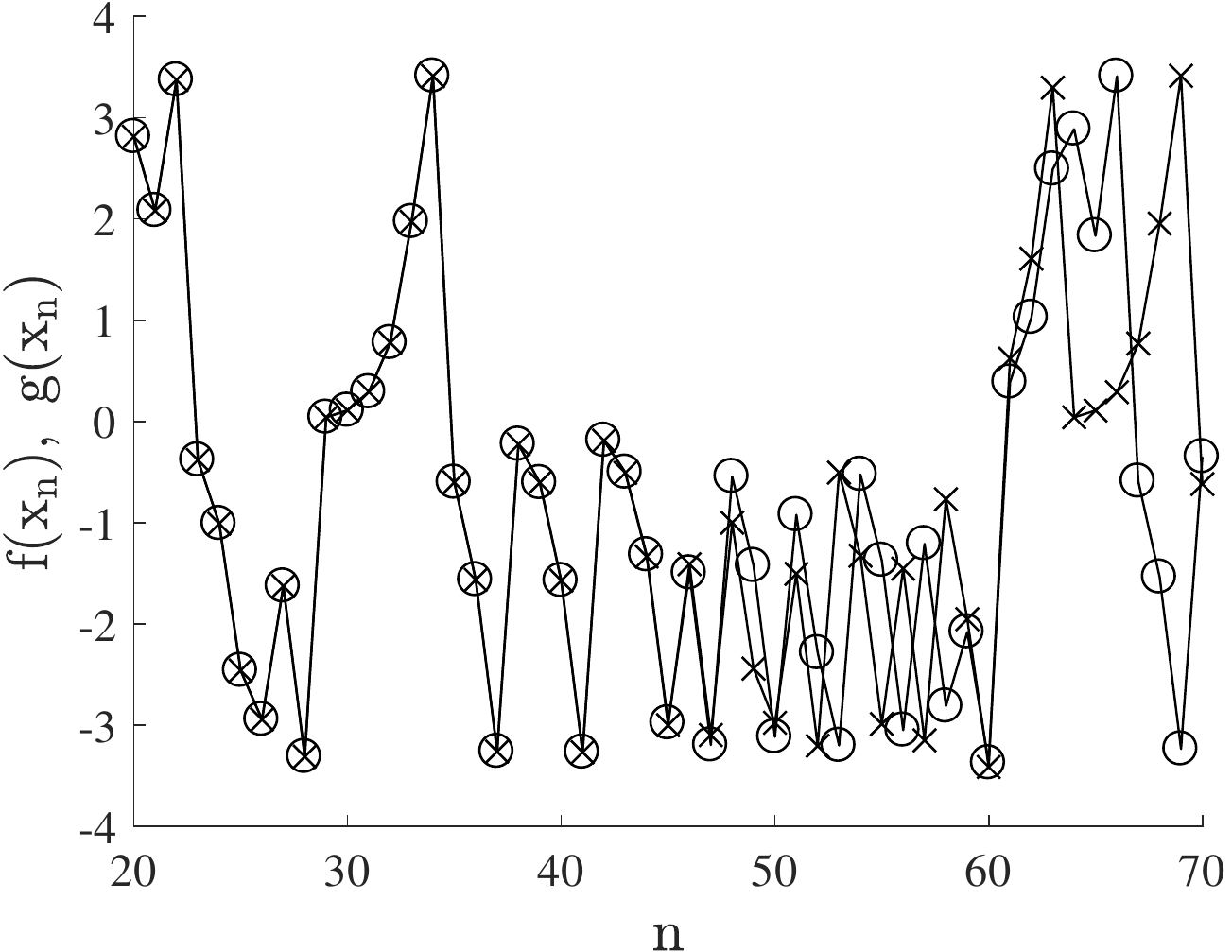}
    \caption{Simulation of Sine Map, with $x_0=0.1$ for two different natural extensions, (-x-) for $f(x_n)$ and (-o-) for $g(x_n)$.}
    \label{sine}
\end{figure}

Some models can be seen as recursive functions in particular non-linear systems \cite{ferreira2006}. Recursive functions are widely used to solve problems and systems, as these functions provide a description for a variety of problems \cite{Feigenbaum1978}. The computational simulations of these systems are subject to errors. To illustrate the error propagation, let us evaluate two natural interval extensions for the Sine Map, identified by the NARMAX model, the first is given by $\mathrm{f(x_n)=2.6868x_n-0.2462x_n^3}$, and the second by its natural extension $\mathrm{g(x_n)=2.6868x_n-(0.2462x_n)x_n^2}$, with $x_0=0.1$. The result obtained is shown in Figure \ref{sine}.

Clearly, the two trajectories are completely different after 50 iterations.  As can be seen by means of  Figure \ref{sine} the propagation of rounding errors may lead to wrong results, even for a very simple system.
A good analysis of the errors arising in numerical calculations can contribute to identify the reliability of calculated results, avoid error risks and improve the accuracy of calculations \cite{Nep2014,Qun2012}.


The use of interval arithmetic \cite{Moore1979} has been considered as an efficient method to deal with numerical errors. The idea is that instead of using a single floating-point value to represent a number, which would result in an error if the number is not representative on the machine, the value is exposed by lower and upper bounds, which define a representable range in the machine \cite{IEEE2015}.  In general, little attention has been given to the propagation of error in the area of system identification. Some works related to this subject was \cite{NM2016,Guedes2017}.
However, in these works the error is investigated when using different extensions. \citeasnoun{NM2016} developed the lower bound error, a practical tool capable of increasing the reliability of the computational simulation of dynamic systems, this lower bound error is specified on basis of two natural extensions. \citeasnoun{Guedes2017} refined the lower bound error, which case it is determined for an arbitrary number of natural extensions. \citeasnoun{PMJN2017} proposed a system identification process focusing on estimation of NARMAX models parameters using the contribution of interval arithmetic with the Intlab toolbox.


The main purpose of this paper is to present an interval bounds using rounding downwards and upwards,  considering the rules of interval arithmetic and basic functions in Matlab.  The method has been applied to Logistic Map \cite{May1976}, the NAR model of the Sine Map \cite{NTAA2003}, and the ARX model of flexible transmission system \cite{LRKV+1995}. To validate the proposed method, the results obtained were compared with the Intlab toolbox.

The rest of the paper is organized as follows. In Section \ref{sec:cp} we recall some preliminary concepts recursive functions and interval arithmetic. Then, in Section \ref{sec:met}, we present the developed method. Section 4 is devoted to present the results, then the final remarks are given in Section 5.

 \section{Preliminary concepts}
 \label{sec:cp}
 \subsection{NARMAX polynomial}

The NARMAX (Non-linear AutoRegressive Moving Average model with eXogenous inputs) model may describe nonlinear systems using difference equations, relating output with linear and nonlinear combinations of past inputs and outputs and can be written as \cite{CB1989} 
\begin{eqnarray}
 y(k) &=& F^l[y(k-1),\cdots ,y(k-n_y),  \nonumber\\
 &&u(k-1),  \cdots , u(k-n_u),   \\ 
&&e(k-1), \cdots , e(k-n_e)]+ e(k), \nonumber 
\end{eqnarray}
where $y(k)$, $u(k)$ e $e(k)$ are, respectively, the output, the input and the noise terms at the discrete time $k \in \mathbb{N}$. The parameters $n_y$, $n_u$ e $n_e$ are their maximum lag. And $F^{\ell}$ is assumed to be a polynomial with nonlinearity degree $\ell$. 


\subsection{Recursive functions}

In recursive functions is possible to calculate the state $x_{n+1}$, at a given time, from an earlier state $x_n$
\begin{equation}
   x_{n+1}=f(x_n), 
\end{equation}
where $f$ is a recursive function and $x_n$ is a function state at the discrete time n. Given an initial condition $x_0$, successive applications of the function $f$ it is possible to know the sequence $\{x_n\} $. The initial condition $x_0$ is called the orbit of $x_0$ \cite{gilmore2012}.

\subsection{Interval Arithmetic}

\citeasnoun{Moore1979} proposed the concept of interval arithmetic, based on the extension of the concept of real numbers to a range of real numbers.

The aim of Moore was to develop a set of techniques capable of producing reliable results, considering rounding errors in the numerical calculation.

 An interval $X$ is denoted as $[\undersl{X},\oversl{X}]$, i.e. $~X ~= ~\{x: ~\undersl{X}\leq ~x\leq ~\oversl{X}\}$, where  $\underline{X}$ and $\overline{X}$, respectively, the lower and upper limit of the interval $X$.  In a degenerated interval, we have $\undersl{X} = \oversl{X}$ and such an interval amounts to a real number $ x = \undersl{X} = \oversl{X}$. 

For a given interval  X =  [$\underline{X}$,$\overline{X}$], its width is defined by $\omega$(X) =( $\overline{X}$ - $\underline{X}$) and its centre is $m(X) =  \frac{1}{2}(\underline{X}+\overline{X}) $ \cite{Rothwell2012}. 

Interval arithmetic provides a method for applying the elementary operations of conventional arithmetic so that the result of the interval includes all possible results. Given $X=[\underline{X},\overline{X}]$ and $Y=[\underline{Y},\overline{Y}]$, the basic interval operations are defined by:

\begin{eqnarray}
X + Y = [\underline{X} + \underline{Y}, \overline{X} +  \overline{Y}],\\
X - Y = [\underline{X} - \overline{Y}, \overline{X} - \underline{Y}],\\
X \cdot Y = [\min\textit{(S)}, \max\textit{(S)}],
\end{eqnarray}
where $S=\{{\underline{X}\underline{Y},\underline{X}\overline{Y}, \overline{X}\underline{Y}, \overline{X}\overline{Y}}$\}. If $0$ does not belong to Y, then $X/Y$ is given by

\begin{equation}
X/Y = X \cdot (1/Y) 
\end{equation}
onde $1/Y = [1/\overline{Y},1/\underline{Y}]$.

\subsection{Intlab}

Intlab is a toolbox for Matlab that supports real and complex intervals, vectors and matrices.  The toolbox was developed so that the computer arithmetic satisfies the IEEE 754 arithmetic standard \cite{ieee754} and, that a permanent switch of the rounding mode is possible. Arithmetical operations in Intlab are accurately ascertained to be correct, comprising input and output and standard functions. By that, it is feasible to supersede every operation of a standard numerical algorithm by the correlating interval operations \cite{Ru99a}.

\section{Methods}
\label{sec:met}


The proposed method is based on the rules of interval arithmetic, so for each operation these rules are obeyed  to obtain a reliable result with the interval arithmetic. In order to achieve the results using intervals, in addition to the arithmetic operations, each operation was performed rounding, so that achieve the lower and upper limits. To illustrate how the simulations were developed, let us consider two intervals $X=[0.1,0.3]$ and $Y=[0.3,0.35]$. Additionally, we perform addition and subtraction operations using the proposed method and the Intlab toolbox. 

\begin{verbatim}
system_dependent(`setround',-Inf)
X_inf=0.1;
Y_inf=0.3;
system_dependent(`setround',Inf)
X_sup=0.3;
Y_sup=0.35;
system_dependent(`setround',-Inf)
X_inf+Y_inf=0.400000000000000
X_inf-Y_sup=-0.250000000000000
system_dependent(`setround',Inf)
X_sup+Y_sup=0.650000000000000
X_sup-Y_inf=-5.551115123125783e-17
X+Y=[0.400000000000000,0.650000000000000]
X-Y=[-0.250000000000000,
-5.551115123125783e-17]
For Intlab toolbox
X=infsup(0.1,0.3);
Y=infsup(0.3,0.35);
X+Y=[0.39999999999998,0.65000000000001]
X-Y=[-0.25000000000000,0.00000000000000] 
\end{verbatim}

 \section{Numerical experiments}

 
 In this section, we present the numerical experiments obtained by the proposed method and the Intlab toolbox. We select three cases studies, which chosen maps are for the systems Logistic Map \cite{May1976}, Sine Map \cite{NTAA2003} and Flexible Transmission Benchmark \cite{LRKV+1995}.
 
 \subsection{Logistic Map}
 
 The logistic map was described by \citeasnoun{May1976} as:
 \begin{equation}
     x_{n+1}=rx_n(1-x_n),
 \end{equation}
 where $r$ is the control parameter, which belongs to the interval $1 \le r \le 4$ and $x_n$ to the interval $0 \le x_n \le 1$. Algorithm 1 explains the simulations performed by the proposed method.

\begin{algorithm}[!ht]
    {\textbf{Algorithm  1} \quad Pseudo-code of the Proposed Method for the Logistic map. \label{alg:1}} \vspace{0.2cm}
      \hrule
    \begin{algorithmic}[1]
        \State \textbf{input} {Number of iterations (N), parameters and initial  conditions}
        \State \verb|system_dependent(`setround',-Inf)|
        \State $\hat{x}_{0,n}^- \leftarrow \hat{x}_{0,n}$
        \State$r^- \leftarrow r$
        \State \verb|system_dependent(`setround',Inf)|
        \State $\hat{x}_{0,n}^+ \leftarrow \hat{x}_{0,n}$
        \State$r^+ \leftarrow r$
        \For{\texttt{n=1:N}}
        \State \verb|system_dependent(`setround',-Inf)|
        \State $\mathrm{aux1_n}   \leftarrow (1-x_{n}^+)$
        \State $\mathrm{aux2_n}   \leftarrow (r^-\times x_{n}^-)$
        \State $x_{n+1}^- \leftarrow \mathrm{aux1_n \times aux2_n}  $
        \State $\mathrm{aux_n} \leftarrow x_{n+1}^-$
         \State \verb|system_dependent(`setround',Inf)|
        \State $\mathrm{aux3_n}   \leftarrow (1-x_{n}^-)$
        \State $\mathrm{aux4_n}   \leftarrow (r^+\times x_{n}^+)$
        \State $x_{n+1}^+ \leftarrow \mathrm{aux3_n \times aux4_n}  $
         \State \verb|system_dependent(`setround',0.5)|
        \If{$x_{n+1}^- >0.5$} 
        \State \verb|system_dependent(`setround',-Inf)|
        \State $x_{n+1}^- \leftarrow x_{n+1}^+$
        \State \verb|system_dependent(`setround',Inf)|
        \State $x_{n+1}^+ \leftarrow \mathrm{aux_{n+1}}$
        \EndIf
        \EndFor
        \State \textbf{output} {$x_{n+1}^-$ and $x_{n+1}^+$}
    \end{algorithmic}
    \hrule
   \end{algorithm}

 \begin{table*}[ht!]
\centering
\caption{Comparison of width size for the simulation of logistic map. The reference is the width produced by means of Intlab (Rump, 1999). We have studied four cases, which the control parameter is $r=3.99$ and initial conditions are as follow: (1): $ x_0 = 0.2 $, (2): $ x_0 = 0.4 $, (3): $x_0 = 0.6$ and (4): $x_0 = 0.8$.}

\label{tab-logi}
\begin{tabular}{c|c|cc|cc}
\hline
\multirow{2}{*}{Case} & \multirow{2}{*}{n} & \multicolumn{2}{c|}{Proposed Method} & \multicolumn{2}{c}{Intlab}       \\ \cline{3-6} 
                      &                    & width  & midpoint   & width & midpoint  \\ \hline
\multirow{4}{*}{(1)}  & 1 & 0 & 0.2& 0 &    0.2           \\
& 5 & 9.7700e-15 & 0.821645072786575& 2.0095e-14              & 0.821645072786575        \\
& 10 & 9.8366e-12
       & 0.973482128268848 & 2.0389e-11&   0.973482128268850          \\
& 20 & 1.0059e-05      & 0.013337715656825 & 2.0851e-05  &  0.013337715672009     \\ \hline
\multirow{4}{*}{(2)}  & 1                  &      0  & 0.4 & 0 & 0.4       \\
& 5 & 1.3212e-14 & 0.990570357273853 & 1.5876e-14  &  0.990570357273853   \\
& 10 & 1.3349e-11 & 0.011714690634153  & 1.6068e-11 & 0.011714690634153 \\
& 20 & 1.3652e-05  & 0.751597796573294 & 1.6432e-05 &    0.751597796578654          \\ \hline
\multirow{4}{*}{(3)}  & 1                  &       0 & 0.6 & 0 &    0.6   \\
& 5 & 1.2768e-14 & 0.990570357273852 & 1.6320e-14 &  0.990570357273852  \\
& 10 & 1.2898e-11 & 0.011714690634148 & 1.6517e-11&  0.011714690634148           \\
& 20 & 1.3190e-05 & 0.751597796556068  & 1.6892e-05  &   0.751597796562074             \\ \hline
\multirow{4}{*}{(4)}  & 1                  &       0 & 0.8 & 0 &   0.8          \\
& 5 & 9.5479e-15 & 0.821645072786574 & 2.0206e-14 &    0.821645072786574         \\
& 10 & 9.6391e-12 & 0.973482128268856 & 2.0502e-11 &  0.973482128268857           \\
& 20 & 9.8575e-06 & 0.013337715653912 & 2.0967e-05  &   0.013337715669766        \\ \hline
\end{tabular}
\end{table*}

 Table \ref{tab-logi} shows the values of a simulation performed in Matlab. This table displays the width and midpoint of the interval, and for comparison shows the results made in the Intlab toolbox.

 \subsection{Sine Map}
 
 A unidimensional sine map is defined as
\begin{equation}
    x_{n+1}=\alpha \sin(x_n),
    \label{eq:sine}
\end{equation}
where $\alpha=1.2\pi$. A polynomial NAR identified for this system is given by \cite{NTAA2003}
\begin{equation}
    y_{n+1}=2.6868y_n-0.2462y^3_n.
    \label{eq:1}
\end{equation}

The  algorithm 2 explains the simulations performed by the proposed method and the  Table \ref{tab-sine} shows the values of a simulation performed in Matlab.

\begin{algorithm}[!ht]
    {\textbf{Algorithm  2} \quad Pseudo-code of the Proposed Method for the Sine map. \label{alg:2}} \vspace{0.2cm}
      \hrule
    \begin{algorithmic}[1]
        \State \textbf{input} {Number of iterations (N) and initial  conditions}
        \State \verb|system_dependent(`setround',-Inf)|
        \State $\hat{x}_{0,n}^- \leftarrow \hat{x}_{0,n}$
        \State \verb|system_dependent(`setround',Inf)|
        \State $\hat{x}_{0,n}^+ \leftarrow \hat{x}_{0,n}$
        \For{\texttt{n=1:N}}
        \State \verb|system_dependent(`setround',-Inf)|
        \State $\mathrm{aux1_n}   \leftarrow 2.6868 \times x_{n}^-$
        \State $\mathrm{aux2_n}   \leftarrow {x_{n}^-}^3$
        \State $\mathrm{aux3_n} \leftarrow 2.2462 \times \mathrm{aux2_n}  $
        \State \verb|system_dependent(`setround',Inf)|
        \State $\mathrm{aux4_n}   \leftarrow 2.6868 \times x_{n}^+$
        \State $\mathrm{aux5_n}   \leftarrow {x_{n}^+}^3$
        \State $\mathrm{aux6_n} \leftarrow 2.2462 \times \mathrm{aux5_n}  $
        \State \verb|system_dependent(`setround',-Inf)|
        \State  $x_{n+1}^- \leftarrow \mathrm{aux1_n} - \mathrm{aux6_n}$
        \State \verb|system_dependent(`setround',Inf)|
      \State  $x_{n+1}^+ \leftarrow \mathrm{aux4_n} - \mathrm{aux3_n}$
        \EndFor
        \State \textbf{output} {$x_{n+1}^-$ and $x_{n+1}^+$}
    \end{algorithmic}
    \hrule
   \end{algorithm}

\begin{table*}[ht!]
\centering
\caption{Comparison of width size for the simulation of sine map. The reference is the width produced by means of Intlab (Rump, 1999). We have studied four cases, which initial conditions are as follow: (1):  $ x_0 = 0.1 $, (2): $ x_0 = 0.2$, (3): $ x_0 = 0.5$, and (4): $x_0=0.8$.}
\label{tab-sine}
\begin{tabular}{c|c|cc|cc}
\hline
\multirow{2}{*}{Case} & \multirow{2}{*}{n} & \multicolumn{2}{c|}{Proposed Method} & \multicolumn{2}{c}{Intlab}       \\ \cline{3-6} 
                      &                    & width  &midpoint  &  width & midpoint  \\ \hline
\multirow{4}{*}{(1)}  & 1 & 0 & 0.1 & 0 &  0.1          \\
& 5 & 1.2879e-14 & 3.408933569627769 & 1.3323e-14              &    3.408933569627769        \\
& 10 & 5.5898e-11
       & -0.847910701541987 & 5.9273e-11 &  -0.847910701542015     \\
& 20 & 1.3298e-04      & 2.811807282224221 & 1.4101e-04 &   2.811807282254640       \\ \hline
\multirow{4}{*}{(2)}  & 1                  &      0  & 0.2 & 0 &   0.2   \\
& 5 & 4.9738e-14 & 1.052283645351666 &  5.7732e-14  &   1.052283645351667   \\
& 10 & 7.3143e-10 & -0.135225347633812  & 8.5036e-10 &   -0.135225347633797 
\\
& 20 & 1.5060e-02   & 0.407773577433717 & 1.7509e-02 &  0.407773930714991          \\ \hline
\multirow{4}{*}{(3)}  & 1                  &       0 & 0.5 & 0 &   0.5     \\
& 5 & 9.5035e-14& 3.229051816564168 & 1.0347e-13 &  3.229051816564168     \\
& 10 & 6.1926e-10 & 1.811036103169470 & 6.7462e-10 &  1.811036103169525    \\
& 20 & 9.7948e-03 & 3.162513358524606  & 1.0670e-02&   3.162513081155066        \\ \hline
\multirow{4}{*}{(4)}  & 1                  &       0 & 0.8 & 0 &   0.8       \\
& 5 & 2.2116e-13 & -1.371443155591735 & 2.6712e-13 &     -1.371443155591733    \\
& 10 & 1.6245e-09
 & -3.378411778526505 & 1.9624e-09 & -3.378411778526536      \\
& 20 & 2.6918e-01 &  0.207611416301658 & 3.2517e-01  &   0.207453930188374            \\ \hline
\end{tabular}
\end{table*}

\subsection{Flexible Transmission Benchmark}

\begin{figure}[ht!]
    \centering
    \includegraphics[scale=0.5]{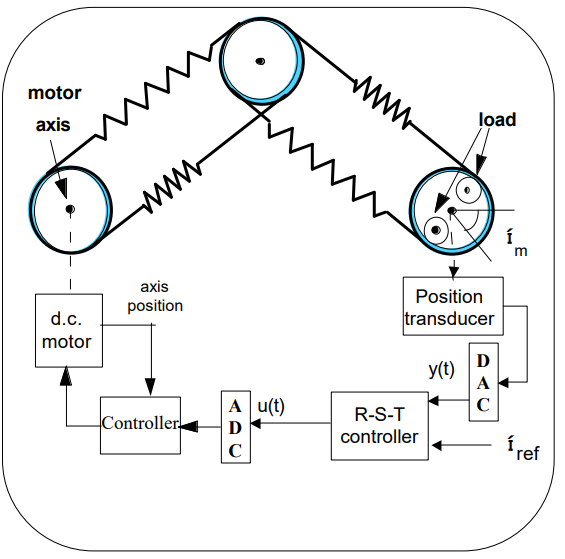}
    \caption{Schematic diagram of the flexible transmission (Landau et al., 1995).}
    \label{fig:l}
\end{figure}

The flexible transmission system of \citeasnoun{LRKV+1995} is presented in Figure \ref{fig:l}. The system input is the reference for the axis position of the first pulley and the system output is the axis position of the third pulley measured by a position sensor, sampled and digitised. The following transfer function is considered as the plant model

\begin{equation*}
    H(q^-1) = \frac{q^{-d}B(q^-1)}{A(q^-1)}.
\end{equation*}

We consider the following ARX model identified and validated  through the real plant \cite{LRKV+1995,BSG2006}:

\begin{small}
\begin{eqnarray}
 y(k) &=&  1.41833y(k-1) - 1.58939y(k-2) + \nonumber\\
 && 1.31608y(k-3) - 0.88642y(k-4)+   \nonumber  \\ 
&& 0.28261u(k-3) + 0.50666u(k-4).
\end{eqnarray}
\end{small}

The pseudo-code of the proposed method for the Flexible Transmission Benchmark is similar to the Sine map. 


\begin{table*}[ht!]
\centering
\caption{Comparison of width size for the simulation of flexible transmission benchmark. The reference is the width produced by means of Intlab (Rump, 1999). We have studied four cases, which initial conditions are as follow: (1):  $ x_0 = 0.1 $, (2): $ x_0 = 0.2$, (3): $ x_0 = 0.6$, and (4): $x_0=0.8$.}
\label{tab-landau}
\begin{tabular}{c|c|cc|cc}
\hline
\multirow{2}{*}{Case} & \multirow{2}{*}{n} & \multicolumn{2}{c|}{Proposed Method} & \multicolumn{2}{c}{Intlab}       \\ \cline{3-6} 
                      &                    & width  &midpoint  &  width & midpoint  \\ \hline
\multirow{4}{*}{(1)}  & 1 & 0 & 0.1 & 0 &  0.1          \\
& 5 & 1.1102e-16 & 0.815130000000000 & 3.3307e-16             &    0.815130000000000        \\
& 10 & 5.4623e-14
       & 1.475024309409214 & 7.5939e-14 &  1.475024309409214    \\
& 20 & 3.2540e-10      & -0.385319792715174 & 4.4843e-10 &   -0.385319792715172
       \\ \hline
\multirow{4}{*}{(2)}  & 1                  &      0  & 0.2 & 0 &   0.2   \\
& 5 & 3.3307e-16 & 0.840990000000000 &  4.4409e-16  &   0.840990000000000  \\
& 10 &6.1062e-14 & 1.432470784456269  & 8.1712e-14 &   1.432470784456269 
\\
& 20 & 3.6389e-10   & -0.406693858836176 & 4.8523e-10 &  -0.406693858836177         \\ \hline
\multirow{4}{*}{(3)}  & 1                  &       0 & 0.6 & 0 &   0.6     \\
& 5 & 5.5511e-16& 0.944430000000000 & 1.1102e-15 &  0.944430000000000     \\
& 10 & 6.5281e-14 & 1.262256684644492 & 1.0791e-13 &  1.262256684644491   \\
& 20 & 3.8598e-10 & -0.492190123320189  & 6.3973e-10&   -0.492190123320189        \\ \hline
\multirow{4}{*}{(4)}  & 1                  &       0 & 0.8 & 0 &   0.8       \\
& 5 & 8.8818e-16 & 0.996150000000000 & 1.4433e-15 &     0.996150000000000    \\
& 10 & 8.3933e-14
 & 1.177149634738603 & 1.1702e-13 & 1.177149634738603      \\
& 20 & 4.9464e-10 &  -0.534938255562194 & 6.9073e-10  &   -0.534938255562196            \\ \hline
\end{tabular}
\end{table*}

In the analysed cases, it was possible to verify that the proposed method which the rules of interval arithmetic are used, there is a decrease of the interval when compared to the Intlab toolbox, since the rounding was performed in each operation and the rounding outward, that is, $\underline{x}$ must be rounded downward and $\overline{x}$ must be rounded upward. And in general, the midpoint average difference is around $6.028e-12$, $9.8825e-6$ and $6.25e-13$, for Logistic Map, Sine Map and Flexible Transmission Benchmark, 
respectively.

\section{Conclusion}

The control of error in numerical simulations has a great importance. Due to computational limitation, the computer does not generate exact answers, but an approximation. Therefore, the interval arithmetic it is possible to say that the result will be within an interval, which this interval encompasses the errors of the simulations, guaranteeing a better evaluation of the result. As previously discussed, there are few publications that report on this subject, mainly using interval arithmetic to examine these errors.


This article has presented a novelty to simulate recursive functions using basic concepts of arithmetic interval and primitive functions in Matlab. The interval calculus of three systems, the logistic map, the sine map, and the flexible transmission benchmark have been performed. To validate this method, the same simulation was executed in the Intlab toolbox. We have found narrower intervals using the proposed method. It is clear that the main advantage of the proposed method is to present a novelty to simulate recursive functions without a specific toolbox. More complex representations for system identification, such as neural networks, should be investigated in future work.

\section*{Acknowledgments}
The authors thank PPGEL, CAPES, CNPq/INERGE, FAPEMIG and the Federal University of São João del-Rei for their support.


\bibliography{cba}
\end{document}